\documentclass{Interspeech2024}

\interspeechcameraready

\title{Optimizing Automatic Speech Assessment: W-RankSim Regularization and Hybrid Feature Fusion Strategies}

\name[affiliation={1}]{Chung-Wen}{Wu}
\name[affiliation={1}]{Berlin}{Chen}

\address{
  $^1$Department of Computer Science and Information Engineering, National Taiwan Normal University}
\email{40947040s@ntnu.edu.tw, berlin@ntnu.edu.tw}

\keywords{automatic speech assessment, ordinal classification, imbalanced data}

\begin{document}

\maketitle

\begin{abstract}
Automatic Speech Assessment (ASA) has seen notable advancements with the utilization of self-supervised features (SSL) in recent research. However, a key challenge in ASA lies in the imbalanced distribution of data, particularly evident in English test datasets. To address this challenge, we approach ASA as an ordinal classification task, introducing Weighted Vectors Ranking Similarity (W-RankSim) as a novel regularization technique. W-RankSim encourages closer proximity of weighted vectors in the output layer for similar classes, implying that feature vectors with similar labels would be gradually nudged closer to each other as they converge towards corresponding weighted vectors. Extensive experimental evaluations confirm the effectiveness of our approach in improving ordinal classification performance for ASA. Furthermore, we propose a hybrid model that combines SSL and handcrafted features, showcasing how the inclusion of handcrafted features enhances performance in an ASA system.
\end{abstract}

\section{Introduction}
Many people learning English as a second language (L2) lack exposure to a native environment and may struggle to receive timely feedback from proficient English speakers. In such scenarios, ASA can readily provide a speech proficiency score, benefiting L2 learners in their practice \cite{eskenazi2009overview, 9414451}. Additionally, it aids language tests by efficiently ensuring more consistent scoring.

Early studies in ASA \cite{zechner2009automatic, qian2019neural, 7404814, 8462562} primarily utilized handcrafted features associated with various aspects of language proficiency, including content (such as relevance), delivery (such as pitch, duration, silence), and language use (such as Part-of-Speech (POS) tags, syntactic dependencies (DEP), morphology), among others. While these features provide interpretability and are directly linked to human grading criteria, they inevitably face challenges in generalizability and are susceptible to errors during handcrafted feature extraction, especially when concerning features related to content and language use. 

Recent research has highlighted the effectiveness of SSL across various speech processing tasks \cite{baevski2020wav2vec, hsu2021robust, yang2021superb, bovbjerg2023improving, xu2021explore, 9206016}, including but is not limited to automatic speech recognition (ASR), keyword spotting, mispronunciation detection and diagnosis (MDD), emotion recognition, and speaker diarisation. Pre-trained models such as wav2vec 2.0 \cite{baevski2020wav2vec}, Whisper \cite{radford2023robust}, and HuBERT \cite{hsu2021hubert} leverage large amounts of data and wherein the contextual information to extract more robust and generalized features. In ASA, these models have demonstrated effectiveness in representing high-level speech-related features and linguistic aspects  \cite{park2023multitask} such as fluency, pronunciation, semantics and even syntax, thereby enhancing overall assessment performance.

A key challenge encountered in ASA is data imbalance, particularly prevalent in datasets sourced from English tests \cite{kim2022automatic, banno2023proficiency}. In these tests, the score distribution tends towards a normal distribution, leading to a scarcity of data points for the lowest and highest scores. Recent research \cite{qian2019neural, park2023multitask, kim2022automatic, banno2023proficiency, park2023addressing} often approaches the ASA task as either a regression or classification problem, utilizing mean square error loss or cross-entropy loss functions for training. However, both of these loss functions neglect the ordinal nature of the scores and are highly affected by data imbalance.

In this paper, we present a hybrid model that combines both pretrained model features and handcrafted features. We conduct a series of experiments and ablation studies to demonstrate how integrating handcrafted features enhances performance. Moreover, we treat the ASA as an imbalanced ordinal classification challenge and leverage the ordinal characteristics of the scores to enhance accuracy. To address this challenge, we put forward an effective optimization framework for ASA modeling, dubbed W-RankSim, which builds upon the concept of RankSim \cite{gong2022ranksim}. 

RankSim is implemented as a batch-wise approach, leveraging ordinal information in regression tasks. However, it often requires a large batch size to demonstrate improvements. In tasks such as ASA, where input signals are substantial, there may not be sufficient VRAM to accommodate such large batch sizes. Additionally, in ordinal classification tasks, RankSim encounters difficulties in accumulating gradients for each class in every batch, as it can only accumulate the gradient of labels in the batch. To address these limitations, we propose W-RankSim, which operates in the weighted vector space. W-RankSim overcomes the batch size constraint of RankSim, reducing training overhead and effectively accumulating gradients for each class in ordinal classification tasks. Furthermore, we identify a suitable loss function to complement W-RankSim. 

In our experiments, we demonstrate that W-RankSim successfully overcomes the limitation of batch size in RankSim and achieves superior performance on ordinal classification tasks.

In summary, this paper presents three main contributions:

\begin{enumerate}
    \item
        Suggesting to approach ASA as an imbalanced ordinal classification problem to enhance the performance by leveraging ordinal information.
    \item 
        Introducing an effective regularization to enhance predictive performance on imbalanced ordinal classification tasks.
    \item 
        Proposing a hybrid model to demonstrate the usefulness of handcrafted features in building an ASA system.
\end{enumerate}
        
To the best of our knowledge, this paper is the first to define ASA as an imbalanced ordinal classification task and propose a pragmatic method to improve performance.

\section{Methodology}
To address the data imbalance problem in ASA systems, which are typically scored using ordinal levels such as CEFR (Common European Framework of Reference for Languages) levels and scores in English tests, we treat the task as an ordinal classification problem. In this context, we proposed W-RankSim (Section \ref{section:W-RankSim}) to leverage the ordinal information in the scores effectively. Additionally, we introduced a novel hybrid model (Section \ref{section:Hybrid model}) to further enhance the performance of ASA systems.

There are some terminologies to clarify: The set of ordinal class labels is denoted by $C$. Our datasets consist of pairs $(x_i, y_i)$, where $x_i$ represents the input vector and $y_i \in C$ denotes the label. The final hidden feature space before predicting the class is denoted as $Z$, where $z_i \in Z$ is obtained by passing $x_i$ through a neural network. The last weight matrix in the output layer is represented by $W \in \mathbb{R}^{|C| \times H^o}$, where $H^o$ is the hidden dimension in output head.

Let $w_j$ denote the weight vector in $W$, where its corresponding label is the $j$-th label. In other words, $w_j$ represents the weight vector associated with the $j$-th class. Consequently, the predicted confidence score of $z_i$ in the $j$-th class is computed as the dot product of $w_j$ and $z_i$, denoted as $w_j \cdot z_i$.

\subsection{W-RankSim}
\label{section:W-RankSim}
Consider an arbitrary vector $a \in \mathbb{R}^n$. Define $\textbf{rk}$ as the ranking function, where $\textbf{rk}(a)$ represents the permutation of $\{1, 2, \dots, n\}$ containing the rank of each element in $a$. In other words, the $i$th element in $\textbf{rk}(a)$ is expressed by
\begin{align}
    \textbf{rk}(a)_i = 1 + |\{j: a_j > a_i\}|
\end{align}
Additionally, let $\sigma^w$ denote the cosine similarity function. We apply cosine similarity to each pair of weighted vectors $(w_i, w_j)$ to obtain the weighted vectors similarity matrix $S^w$. Each entry in $S^w \in \mathbb{R}^{|C|\times|C|}$ can be represented by
\begin{align}
    S^w_{i,j} = \sigma^w(w_i, w_j)
\end{align}
In the label space, each pair of class labels $(c_i, c_j)$, where $c_i, c_j \in C$, are passed to $\sigma^c$,which is a simple negative absolute distance function, to construct $S^c \in \mathbb{R}^{|C|\times|C|}$:
\begin{align}
    S^c_{i,j} = \sigma^c(c_i, c_j)
\end{align}
Then, W-RankSim in ordinal class labels $C$ is constructed as:
\begin{align}
\label{equation:eq4}
L_{\text{W-RankSim}} = \sum^{|C|}_{i=1} l\left(\textbf{rk}(S^c_{[i,:]}), \textbf{rk}(S^w_{[i,:]})\right)
\end{align}
where $[i,:]$ denotes the $i$-th column in the matrix, and $l$ is a ranking similarity function that aims to make the rank similarity between the label space and weighted vector space similar. In this case, mean square error is used as $l$. 

During training, the main loss function $L_{\text{main}}$ is combined with W-RankSim. The complete formula is as follows:
\begin{align}
L_{\text{total}} = L_{\text{main}} + \gamma L_{\text{W-RankSim}}
\end{align}
where $\gamma$ is a hyperparameter that controls the magnitude of the W-RankSim regularizer. W-RankSim encourages weight vectors with closer class labels to be closer in cosine space. Furthermore, $L_{\text{main}}$ ensures that feature vectors $z_i$ are centered at their corresponding weighted vectors $w_j$. As $z_i$ approaches the corresponding $w_j$, it simultaneously moves closer to other feature vectors with closer labels and moves away from those with farther labels.

In optimization, the $\textbf{rk}$ function in Eq. \ref{equation:eq4} is piece-wise constant and non-differentiable. We adopt the method proposed in \cite{rolinek2020optimizing} to reformulate the $\textbf{rk}$ function as a combinatorial objective:
\begin{align}
\textbf{rk}(a) = \arg\min_{\pi \in \prod_n} a \cdot \pi
\end{align}
where $\prod_n$ is a set containing all permutations of $\{1, 2, \dots,n\}$. This formulation enables us to utilize blackbox combinatorial solvers \cite{poganvcic2019differentiation} for differentiation.

The update formula is parameterized by a hyperparameter $\lambda$, which balances the faithfulness to the original function and the informativeness of the gradient, and is expressed by
\begin{align}
\frac{\partial L}{\partial a} = -\frac{1}{\lambda}(\textbf{rk}(a) - \textbf{rk}(a_{\lambda})), \
a_{\lambda} = a + \lambda \frac{\partial L}{\partial \textbf{rk}}
\end{align}

For further details, please refer to \cite{rolinek2020optimizing, poganvcic2019differentiation}.

\subsection{Comparing with RankSim}
\begin{figure}[t]
  \centering
  \includegraphics[width=\linewidth]{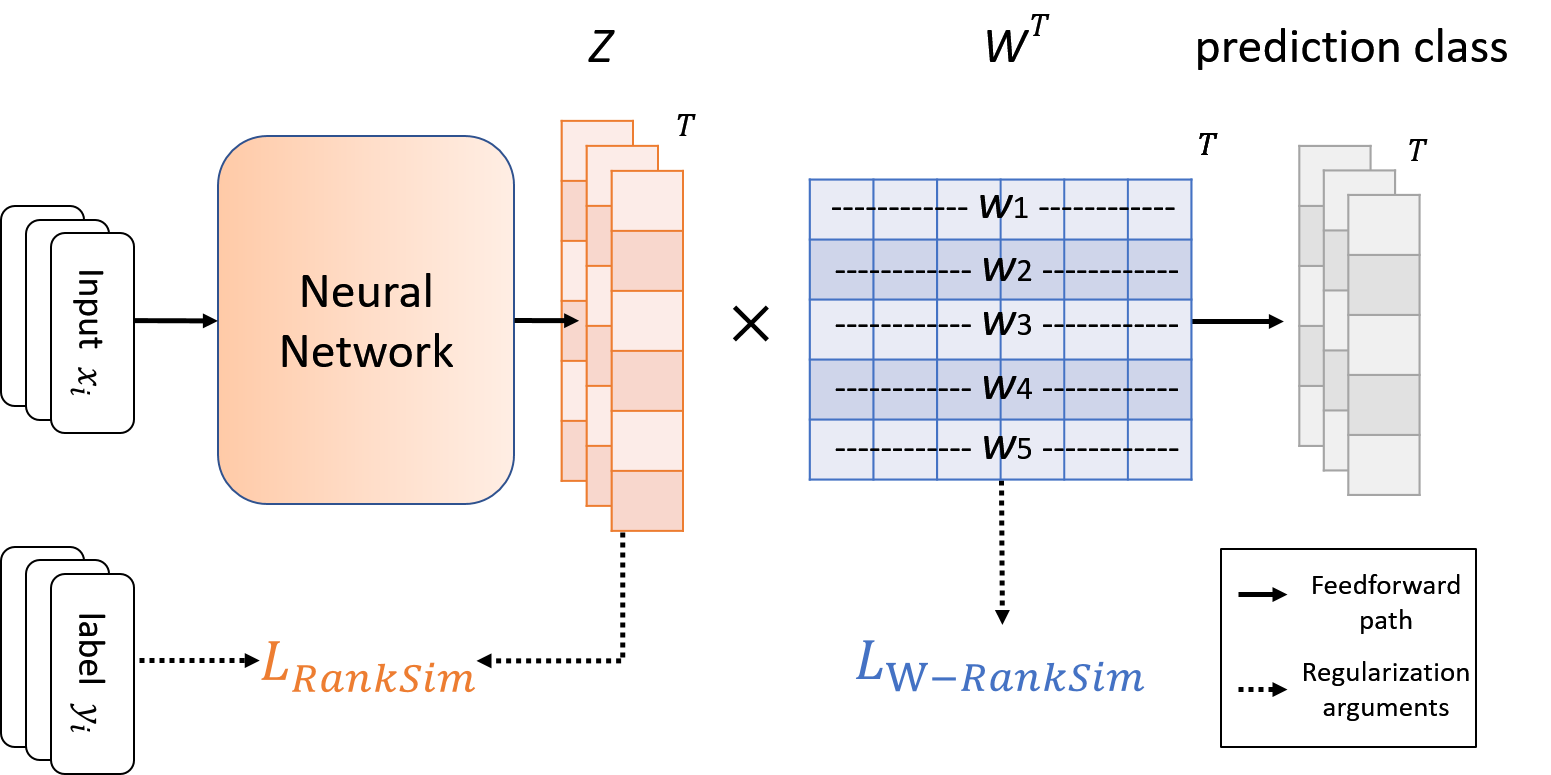}
  \caption{Illustration of our proposed regularization W-RankSim vs. RankSim. RankSim requires the last feature embeddings and labels in a batch, whereas W-RankSim leverages weighted vectors in the output layer without being constrained on batch size to achieve a similar effect.}
  \label{fig:compare}
\end{figure}

As illustrated in Figure \ref{fig:compare}, our proposed method directly imposes constraints on the weighted vectors, ensuring that ordinal information between each class is considered at each updating step, regardless of whether there is data imbalance. In contrast, RankSim is confined to the samples within the batch. Although it employs a subsampling technique to ensure that each label is considered only once in a batch, it still encounters inherent limitations, particularly in imbalanced ordinal classification tasks where the number of labels is typically small. In such cases, a large enough batch size is needed to ensure that  the labels in a batch can offer sufficient variation and information. However, in our ASA task, a large batch size is not feasible due to the large input signal. In summary, W-RankSim is an approach that is not constrained by batch size and achieves, or even surpasses, the effectiveness of RankSim with lower overhead in ordinal classification tasks.

\subsection{Hybrid model}
\label{section:Hybrid model}
\begin{figure*}[t]
  \centering
  \includegraphics[width=\linewidth]{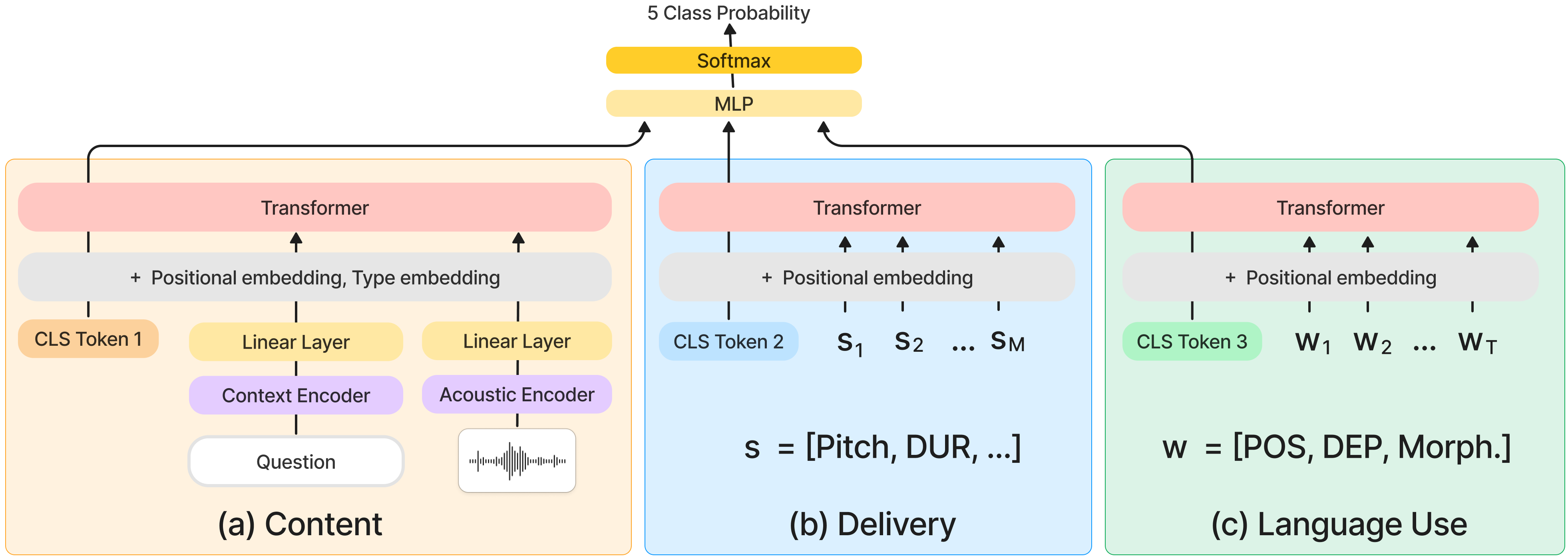}
  \caption{The architecture of the Hybrid ASA model comprises three parts: (a) content, (b) delivery, and (c) language use, each addressing specific aspects of speech assessment. The first part utilizes SSL features generated by a pretrained acoustic model such as Whisper. The remaining parts leverage hand-crafted features to capture relevant characteristics for speech assessment.}
  \label{fig:hybrid model}
  \vspace{-0.5cm}
\end{figure*}

Figure \ref{fig:hybrid model} illustrates our proposed hybrid model, which integrates SSL features with handcrafted features. The model consists of three main components: (a) content, (b) delivery, and (c) language use.

\subsubsection{Content component}
This component aims to evaluate the relevance between the user response and the question. We begin by encoding the audio signal and question context separately using pretrained context encoder and acoustic encoder, specifically BERT and Whisper, to obtain sequence embeddings that represent the questions and audio sequences while preserving contextual information. Subsequently, these embeddings are projected into the same space with $H$ dimension using a linear layer. To distinguish between the embeddings of the two modalities (text and audio) and to capture the temporal sequence characteristics, we introduce learnable type embeddings and positional embeddings. Finally, we employ the CLS token by concatenating it with all embeddings. This concatenated representation is then passed through bidirectional transformer encoder \cite{vaswani2017attention} blocks to learn vector representations of the content features.

\subsubsection{Delivery component}
This aspect measures users' stress, pronunciation, fluency, etc. In this component, we utilize monolingual wav2vec2 2.0 for force alignment with ASR transcriptions, recognized by multilingual wav2vec 2.0, to segment the audio signal into word-level segments using the CTC segmentation algorithm \cite{kurzinger2020ctc}. Within each segmentation, we extract various features including pitch, duration (DUR), intensity, following silence, posterior probability from both monolingual and multilingual wav2vec 2.0, LM score from n-gram LM during transcript, and confidence score from multilingual wav2vec 2.0. These features are utilized to construct segment features $s_i$, and subsequently, a sequence of segment features $[s_1, s_2, \dots, s_M]$ is concatenated to represent the delivery feature, where $M$ is a predefined length achieved by either truncating or padding with zeros. To learn vector representations of these delivery features, we employ the same transformer encoder blocks, learnable positional embeddings, and CLS token as utilized in the content component.

\subsubsection{Language use component}
Language use measures users' abilities in grammar and syntax. In this component, we obtain a word sequence from the ASR transcription generated by the delivery component and adjust it to length $T$ by either truncating or padding with zeros. For each word, we utilize spaCy, an industrial-strength natural language processing toolkit in Python, to extract Part-Of-Speech (POS), dependency labels (DEP), and morphology (Morph.). These features are then used to construct a language use embedding $w_i$ of a word using one-hot encoding. Subsequently, a sequence of embeddings is concatenated $[w_1, w_2, \dots, w_T]$ to represent the sequence. Finally, we adapt the same steps as in the delivery component, using the CLS token to represent the entire sequence.

These three components form the foundation for comprehensive evaluation \cite{qian2019neural, park2023multitask}. Similar to the approach proposed in \cite{park2023addressing}, we incorporate prompting to adapt to various questions. Finally, the CLS token embeddings obtained from these three components are concatenated and passed through a multilayer perceptron (MLP) with $H^o$ hidden dimensions. We apply the softmax function to the output to obtain the probability distribution over $|C|$ classes. In our case, $|C|$ is $5$.

\section{Experiments and Results}
\subsection{Data}

\vspace{-0.5cm}

\begin{figure}[h]
  \centering
  \includegraphics[width=\linewidth]{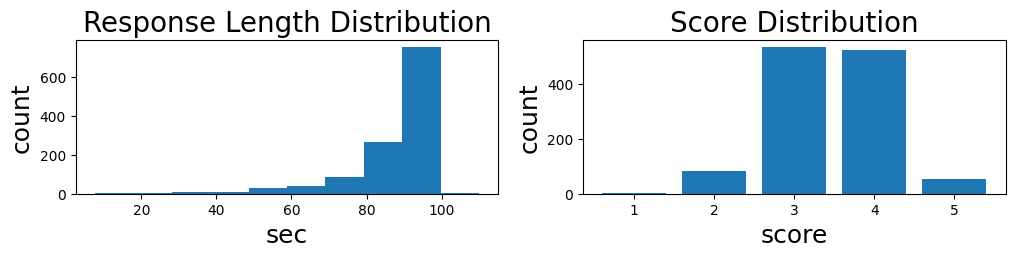}
  \vspace{-0.7cm}
  \caption{The left figure shows response length distribution in the GEPT corpus, while the other displays score distribution.}
  \vspace{-0.3cm}
  \label{fig:data}
\end{figure}


The corpus used in this study was privately collected by the Language Training and Testing Center from the General English Proficiency Test (GEPT) intermediate level exam, specifically focusing on the picture description module. This exam is a high-stakes English assessment test. The corpus comprises 1199 responses evenly distributed across 4 sets of questions, each provided by a different test taker. Participants were instructed as follows: "Below are a picture and four related questions. Please complete your answers in one and a half minutes. Do not read the number or the question when you answer. Please first look at the picture and think about the questions for thirty seconds."

The score distribution and response length distribution are illustrated in Figure \ref{fig:data}. Scores were calculated as averages from assessments provided by two professionals, ranging from 1 to 5, and any floating-point values were discarded unconditionally.

We randomly selected a specific set of questions to create an unknown content test set. This set was designed to assess performance under cold start conditions \cite{park2023addressing}, where the model encounters previously unseen content. The remaining data was divided into train, development, and known content test sets in an 8:1:1 ratio. This partitioning strategy ensures that the model is trained on a majority of the data while allowing for robust evaluation on both familiar and novel content.

\subsection{Experimental Setup}
We employed Whisper-base as our acoustic encoder and utilized Sentence-BERT \cite{reimers2019sentence}, specifically the all-MiniLM-L6-v2 model, as our context encoder. The dimensions for both the hidden layers ($H$) and the hidden layers in the MLP ($H^o$) were set to 512 and 64, respectively. During training, we utilized the RAdam optimizer \cite{liu2019variance} with a weighted decay of 1e-5. The learning rate was set to 2e-4, and training was conducted for 8 epochs with a batch size of 2.

The experiments were conducted with W-RankSim and RankSim using both cross-entropy loss and large margin cosine loss (LMCL) \cite{wang2018cosface}. We set the hyperparameters $\lambda$ and $\gamma$ to 2 and 1.5, respectively, for W-RankSim and RankSim. LMCL has two hyperparameters, $s$ and $m$, which were set as 1.96 and 0.15 in all experiments. Further details about $s$ and $m$ can be found in \cite{wang2018cosface}. This configuration allowed us to effectively train our model while optimizing performance.

\subsection{Performance comparison}

\vspace{-0.3cm}

\begin{table}[h]
\resizebox{\columnwidth}{!}{%
\begin{tabular}{@{}lcc@{}}
\toprule
Hybrid model                            & \begin{tabular}[c]{@{}c@{}}Known Content\\ Accuracy \end{tabular} & \begin{tabular}[c]{@{}c@{}}Unknown Content\\ Accuracy\end{tabular}        \\ \midrule
cross entropy loss                      & 0.689          & 0.620          \\
+W-RankSim                              & 0.678          & 0.700          \\
LMCL                                    & 0.678          & 0.687          \\
+W-RankSim                              & \textbf{0.700}   & \textbf{0.720}  \\ \midrule
w/o language use component              & \begin{tabular}[c]{@{}c@{}}Known Content\\ Accuracy \end{tabular} & \begin{tabular}[c]{@{}c@{}}Unknown Content\\ Accuracy\end{tabular}        \\ \midrule
cross entropy loss                      & 0.633          & 0.683          \\
+W-RankSim                              & 0.644          & 0.700          \\
LMCL                                    & 0.644          & 0.700          \\
+W-RankSim                              & \textbf{0.656} & \textbf{0.717} \\ \midrule
w/o delivery and language use component & \begin{tabular}[c]{@{}c@{}}Known Content\\ Accuracy \end{tabular} & \begin{tabular}[c]{@{}c@{}}Unknown Content\\ Accuracy\end{tabular}        \\ \midrule
cross entropy loss                      & 0.633          & 0.657          \\
+W-RankSim                              & 0.644          & \textbf{0.697} \\
LMCL                                    & 0.644          & 0.680          \\
+W-RankSim                              & \textbf{0.678} & 0.683          \\ \bottomrule
\end{tabular}%
}

\caption{Experiment results on the GEPT corpus. "Known Content Accuracy" denotes the accuracy on the known content test set, while "Unknown Content Accuracy" represents the accuracy on the unknown content test set.}
\label{table:experiment}
\end{table}
 
\vspace{-0.5cm}

In Table \ref{table:experiment}, we conduct an ablation study on model components and demonstrate the effectiveness of W-RankSim in each model. We also experimented with various loss functions, including cross-entropy loss and LMCL. Across each model and loss function, incorporating W-RankSim consistently improved performance, indicating its benefit in ordinal classification tasks with imbalanced data.

The experiments demonstrate that both cross-entropy loss and LMCL combined with W-RankSim have improved performance in known content test set and unknown content test set, particularly with the inclusion of LMCL. LMCL in conjunction with W-RankSim consistently yields better performance in most models compared to using cross-entropy with W-RankSim. According to our observations, LMCL enhances local relations in cosine space, while W-Ranksim emphasizes global relations between each class in cosine space, thus improving performance. This suggests that leveraging these methods together enhances the classification accuracy.

Our baseline model consists solely of the content component. The experiments presented in Table \ref{table:experiment} demonstrate the advantages of integrating handcrafted features. Both language use and delivery features contribute to performance improvement, with the fully hybrid model achieving the best performance on both the known content test set and the unknown content test set

\subsection{Experiments on varying batch size.}
\begin{figure}[h]
  \centering
  \vspace{-0.2cm}
  \includegraphics[width=\linewidth]{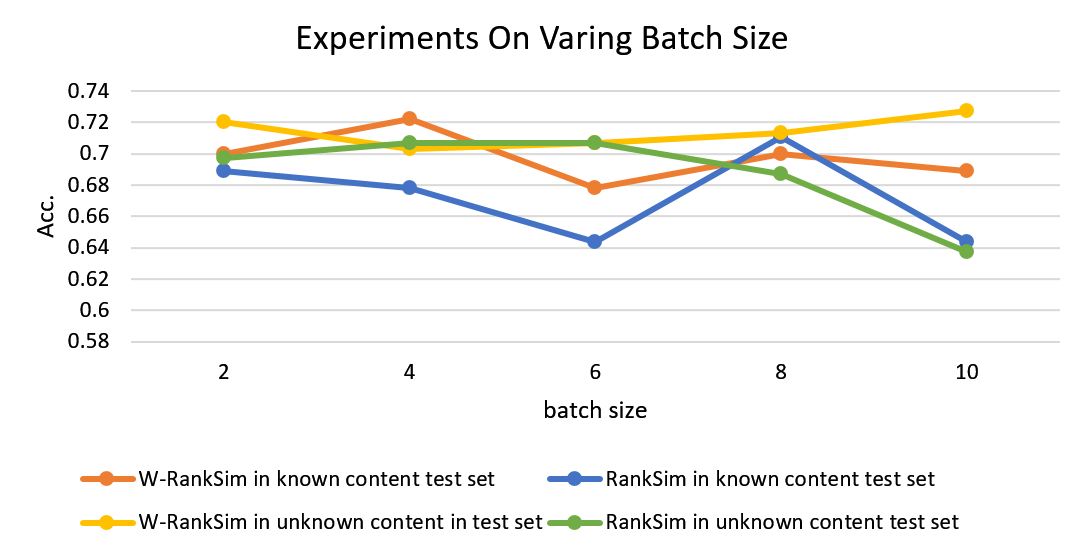}
  \caption{Experiments tested a hybrid model that combined LMCL with different regularization across different batch sizes.}
  \label{fig:batch size}
\end{figure}

\vspace{-0.3cm}

We conducted experiments varying the batch size using the hybrid model with LMCL combined with W-RankSim and RankSim to assess the sensitivity of batch size in W-RankSim and RankSim. The results depicted in Figure \ref{fig:batch size} demonstrate that performance in W-RankSim is more consistent compared to RankSim when changing the batch size, and W-RankSim can achieve the best or near-best performance at lower batch size. 

\section{Conclusions}
We approached the ASA task as an ordinal classification task with imbalanced data and introduced a novel regularization method, W-RankSim, specifically designed for this task. W-RankSim captures ordinal information between weighted vectors, indirectly encouraging embeddings to learn proximity and distance relations in both label and feature space. Our experiments consistently demonstrate that W-RankSim can improve performance across various models. Subsequently, we proposed a hybrid model that combines SSL features with traditional hand-crafted features and demonstrated its effectiveness through extensive experiments. Finally, the hybrid model using LMCL with W-RankSim achieved the best accuracy and exhibited steady performance across varying batch sizes. This underscores the robustness and efficacy of the proposed approach.

\textbf{Limitations and future work.} In this paper, we focus on exploring regularization techniques and feature extraction methods to enhance ASA systems. However, we acknowledge the significance of linguistic and phonetic aspects in real-world speech assessment, which were not fully addressed in our work. Future effort will involve collaborating closely with linguistics and phonetics experts to develop a more comprehensive ASA system that integrates these crucial factors. Additionally, we aim to investigate the applicability of W-RankSim in other ordinal classification tasks to advance classification techniques further and explore its potential in broader contexts.

\section{Acknowledgement}
This work was supported by the Language Training and Testing Center (LTTC), Taiwan. Any findings and implications in the paper do not necessarily reflect those of the sponsor.

\bibliographystyle{IEEEtran}
\bibliography{mybib}

\end{document}